\documentclass{kluwer}    % Specifies the document style.

\input psfig.tex

\begin{document}                                                                                   
%Luis's definitions
\def\aa{{A\&A}}
\def\aas{{ A\&AS}}
\def\aj{{AJ}}
\def\al{$\alpha$}
\def\bet{$\beta$}
\def\amin{$^\prime$}
\def\annrev{{ARA\&A}}
\def\apj{{ApJ}}
\def\apjs{{ApJS}}
\def\asec{$^{\prime\prime}$}
\def\baas{{BAAS}}
\def\cc{cm$^{-3}$}
\def\deg{$^{\circ}$}
\def\ddeg{{\rlap.}$^{\circ}$}
\def\dsec{{\rlap.}$^{\prime\prime}$}
\def\cc{cm$^{-3}$}
\def\e#1{$\times$10$^{#1}$}
\def\etal{{et al. }}
\def\flamb{erg s$^{-1}$ cm$^{-2}$ \AA$^{-1}$}
\def\flux{erg s$^{-1}$ cm$^{-2}$}
\def\fnu{erg s$^{-1}$ cm$^{-2}$ Hz$^{-1}$}
\def\hal{H$\alpha$}
\def\hst{{\it HST}}
\def\kms{km s$^{-1}$}
\def\lamb{$\lambda$}
\def\lax{{$\mathrel{\hbox{\rlap{\hbox{\lower4pt\hbox{$\sim$}}}\hbox{$<$}}}$}}
\def\gax{{$\mathrel{\hbox{\rlap{\hbox{\lower4pt\hbox{$\sim$}}}\hbox{$>$}}}$}}
\def\simlt{\lower.5ex\hbox{$\; \buildrel < \over \sim \;$}}
\def\simgt{\lower.5ex\hbox{$\; \buildrel > \over \sim \;$}}
\def\lum{erg s$^{-1}$}
\def\mbh{{$M_{\rm BH}$}}
\def\micron{{$\mu$m}}
\def\mnras{{MNRAS}}
\def\nat{{Nature}}
\def\pasp{{PASP}}
\def\perang{\AA$^{-1}$}
\def\percm2{cm$^{-2}$}
\def\pp{\parshape 2 0truein 6.1truein .3truein 5.5truein}
\def\refindent{\par\noindent\parskip=2pt\hangindent=3pc\hangafter=1 }
\def\solum{$L_\odot$}
\def\solmass{$M_\odot$}
\def\ion#1#2{\setcounter{ctr}{#2}#1$\;${\small\rm\Roman{ctr}}\relax}
\def\oii{[\ion{O}{2}]}
\def\heii{\ion{He}{2}}
\def\hi{\ion{H}{1}}
\def\hii{\ion{H}{2}}
\def\oiii{[\ion{O}{3}]}
\def\ni{[\ion{N}{1}]}
\def\oi{[\ion{O}{1}]}
\def\nii{[\ion{N}{2}]}
\def\hei{\ion{He}{1}}
\def\sii{[\ion{S}{2}]}
\def\siii{[\ion{S}{3}]}

%\newdisplay{guess}{Conjecture}

\begin{article}         
\begin{opening}         
\title{``Low-state'' Black Hole Accretion in Nearby Galaxies\footnote{
To appear in {\it From X-ray Binaries to Quasars: Black Hole Accretion on
All Mass Scales}, ed. T. J. Maccarone, R. P. Fender, and L. C. Ho
(Dordrecht: Kluwer)}}
\author{Luis C. \surname{Ho}}  
\runningauthor{Luis C. Ho}
\runningtitle{``Low-state'' Black Hole Accretion}
\institute{The Observatories of the Carnegie Institution of Washington, 
813 Santa Barbara Street, Pasadena, CA 91101, U.S.A.}

\begin{abstract}
I summarize the main observational properties of low-luminosity AGNs in nearby 
galaxies to argue that they are the high-mass analogs of black hole X-ray 
binaries in the ``low/hard'' state.  The principal characteristics of 
low-state AGNs can be accommodated with a scenario in which the central engine 
is comprised of three components: an optically thick, geometrically 
accretion disk with a truncated inner radius, a radiatively inefficient 
flow, and a compact jet.
\end{abstract}

\end{opening}           

\section{AGNs and Black Hole X-ray Binaries}

Stellar-mass black holes in X-ray binaries, in 
response to changes in the mass accretion rate, exhibit 
distinct spectral ``states'' (McClintock \& Remillard 2005).  Since many 
aspects of accretion flows are invariant with respect to changes in black 
hole mass, it is of interest to ask whether there are extragalactic analogs to 
X-ray binary states in massive black holes in the centers of galaxies.  Nuclear 
black holes outweigh stellar black holes by factors of $10^5-10^8$, and so 
their evolutionary timescales increase in the same proportion.  To search for 
spectral states in massive black holes, one must consider the demographics of 
accreting nuclear black holes---AGNs---spanning a wide range of luminosity. 

In recent years, much attention has been devoted to the study of 
``narrow-line'' Seyfert 1 galaxies, which are widely believed to be 
the AGN counterparts of X-ray binaries in the ``high/soft'' state 
(e.g., Pounds, Done, \& Osborne 1995).  Indeed, this class of AGNs may be even 
accreting at super-Eddington rates (Collin \& Kawaguchi 2004). 

This contribution focuses on AGNs in the opposite extreme, namely those 
accreting at highly sub-Eddington rates, which I will argue are close analogs
to X-ray binaries in the ``low/hard'' or ``quiescent'' states, and which 
dominate the population of AGNs at $z = 0$.

\section{Observational Properties of Low-luminosity AGNs}

Although the physical nature of low-luminosity AGNs (LLAGNs) is still not
fully understood, the bulk of the current evidence suggests that a significant 
fraction of them are genuinely accretion-powered sources (for a recent review, 
see Ho 2004).  Here I highlight the most important observational properties of 
these objects, which, when taken collectively, point to some novel insights 
on the structure of their central engines.

\begin{itemize}

\item{
{\it Demography.}\ LLAGNs are very common.  According to the Palomar
survey (Ho, Filippenko, \& Sargent 1997a), over 40\% of nearby galaxies, and an
even greater fraction (50\%--75\%) of bulge-dominated (E--Sbc) systems, contain 
LLAGNs.
}

\item{
{\it Low ionization.}\ The dominant population (2/3) of LLAGNs have
low-ionization state spectra (Ho et al. 1997a).  They are classified as either 
LINERs (low-ionization nuclear emission-line regions) or transition objects, 
which are hypothesized to be related to LINERs (Ho, Filippenko, \& Sargent 
1993; but see complications discussed in Ho, Filippenko, \& Sargent 2003 and 
Ho 2004).
}

\item{
{\it Low accretion power.}\  LLAGNs are intrinsically faint, in most 
cases orders of magnitude less powerful than classical Seyferts and quasars.
The optical luminosity function of LLAGNs extends to absolute magnitudes 
as low as $M_B \approx -6$ (Ho 2004).  Figure~1{\it a}, from Ho (2005), shows 
the distributions of bolometric luminosities for $\sim$250 objects from the 
Palomar survey.  Note that nearly all the objects have $L_{\rm bol} < 10^{44}$ 
\lum, and most significantly less. Seyferts are on average 10 times more 
luminous than LINERs or transition objects.
}

\item{
{\it Sub-Eddington.}\  LLAGNs are highly sub-Eddington systems, as shown
in Figure~1{\it b}.  Essentially all objects in the Palomar sample 
have $L_{\rm bol}/L_{\rm Edd} < 1$, with the majority falling in the 
region $L_{\rm bol}/L_{\rm Edd} \approx 10^{-5}-10^{-3}$.
Seyferts have systematically higher Eddington ratios than LINERs or
transition objects, typically by 1 to 2 orders of magnitude.
}

\item{
{\it Radiatively inefficient.}\  Direct measurements of accretion rates are
not available, but rough estimates can be made of the likely minimum rates 
supplied {\it in situ}.  Quite apart from any additional fuel furnished
by the large-scale disk of the host galaxy or from external sources 
triggered by tidal interactions, both of which are inconsequential for 
nearby galaxies (Ho et al. 1997b, 2003), galactic centers contain two 
reservoirs of fuel that seems inescapable.   The first is mass loss from 
evolved stars, which is readily available from the dense stellar cusps 
invariably observed in the centers of bulges (e.g., Lauer et al. 1995; 
Ravindranath et al. 2001).  Ho (2005) estimates 
$\dot M_*$ \gax\ $10^{-5} - 10^{-3}$ \solmass\ yr$^{-1}$.  The other source 
of fuel is Bondi accretion of hot gas, which is ubiquitous not only in 
giant elliptical galaxies, but apparently also in the bulges of disk galaxies 
(e.g., Shirey et al. 2001; Baganoff et al. 2003).  The expected contribution 
from Bondi accretion turns out to be roughly comparable to $\dot M_*$ (Ho 
2005).  If this gas were to be all accreted and radiates with a standard
efficiency of $\eta = 10$\%, the nuclei should be $1-4$ orders of magnitude
more luminous than observed.  Three possible explanations come to mind: (1) 
angular momentum transfer is very inefficient, even at these small scales, so 
that only a tiny fraction of the available fuel makes it to the center; 
(2) the accretion flow is radiatively inefficient, with $\eta$ much less than 
10\%; (3) most of the gas is blown out of the system by winds or outflows, 
which arise naturally in radiatively (e.g., Blandford \& Begelman 1999).  
(Note that the third option is not entirely independent from the second.) 
While it is difficult to rule out the first explanation, it seems plausible 
that these systems are radiatively inefficient.
}

%Figure 1
%%BoundingBox: 50 50 440 840
\begin{figure}
\vbox{
\hbox{
\hskip -0.3truein
\psfig{file=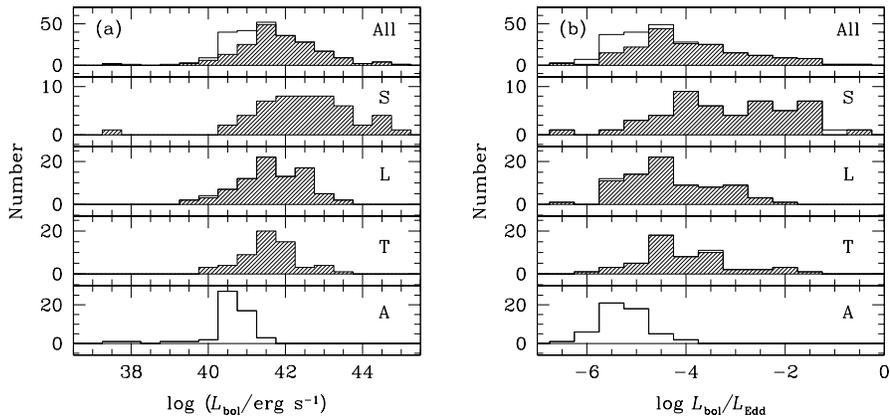,height=2.5truein,angle=270}
}
}
\caption{
Distribution of ({\it a}) nuclear bolometric luminosities
and ({\it b}) Eddington ratios, $L_{\rm bol}/L_{\rm Edd}$.
S = Seyferts, L = LINERs, T = transition objects, and A = absorption-line
nuclei.  Open histograms denote upper limits.  From Ho (2005.)}
\end{figure}

\item{
{\it Unusual SEDs.}\   With few exceptions, the spectral energy
distributions (SEDs) of LLAGNs lack the optical--UV ``big blue bump,'' a
feature usually attributed to thermal emission from an optically thick,
geometrically thin accretion disk (Ho 1999, 2002a; Ho et al. 2000).  This is 
illustrated in Figure 2, which compares the average SED of LLAGNs with the 
canonical SEDs of radio-loud and radio-quiet quasars (Elvis et al. 1994).
Instead of a blue excess, there is a maximum peaking somewhere in the mid-IR.
(The exact location of the peak is poorly defined because of the current 
lack of high-resolution IR data.)  One consequence of the deficit of 
optical--UV omission is that the X-rays become disproportionately 
important energetically.  The standard $\alpha_{\rm ox}$ parameter is 
typically less than 1, whereas in luminous AGNs $\alpha_{\rm ox} \approx 1.4$.  
The X-ray spectra can be well described by a
simple power law, with $\Gamma \approx 1.7-1.9$, which generally requires 
only little or modest intrinsic absorption, with no evidence for a 
soft excess at low energies (e.g., Terashima et al. 2002; Terashima \& 
Wilson 2003; Ptak et al. 2004).
}

%Figure 2
\begin{figure}
\vbox{
\hbox{
\hskip 0truein
\psfig{file=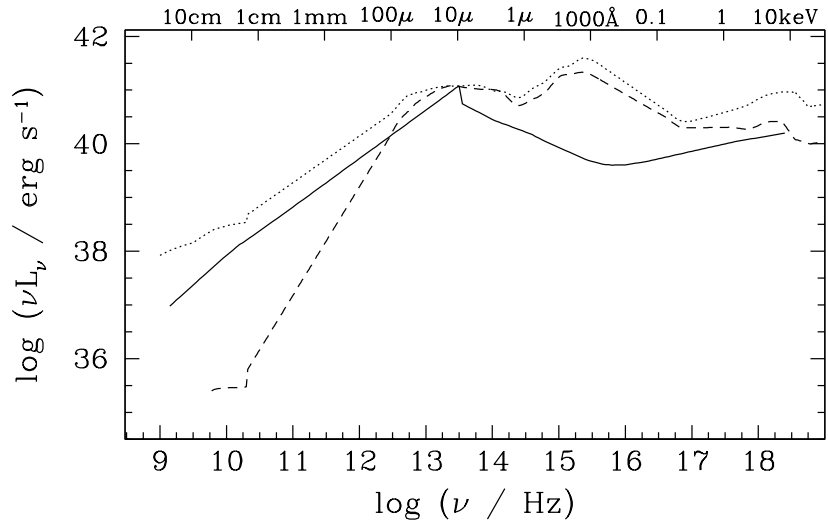,width=5.0truein,angle=270}
}
}
\caption{
The average SED of low-luminosity AGNs ({\it solid line}), adapted from
Ho (1999).  Overplotted for comparison are the average SEDs of powerful
radio-loud ({\it dotted line}) and radio-quiet ({\it dashed line}) AGNs (Elvis
\etal 1994).  The curves have been arbitrarily normalized to
the luminosity at 10 $\mu$m.}
\end{figure}

\item{
{\it Radio jets.}\   Another notable feature of the SEDs of LLAGNs is that they 
tend to be generically radio-loud.  This is true of most LINERs (Ho 1999, 2002b; 
Ho et al.  2000; Terashima \& Wilson 2003), and, contrary to persistent popular
misconception, is so even in most Seyfert nuclei (Ho \& Peng 2001).  Detailed 
modeling of the SEDs (e.g., Quataert et al. 1999; Ulvestad \& Ho 2001; 
Anderson, Ulvestad, \& Ho 2004) shows that neither the radio power nor the 
detailed radio spectrum agrees with predictions from accretion flow models. 
Instead, a separate, compact jet component is required.  This indicates that 
compact jets develop naturally in LLAGNs.
}

\item{
{\it No broad Fe K\al\ line.}\  The 6.4 keV Fe K\al\ line is detected in
some LLAGNs, but it is almost always narrow (Terashima et al. 2002).  In 
well-studied cases (e.g., Ptak et al. 2004), Fe K\al\ emission of any 
breadth can be ruled out to very high significance.  Insofar as the broad iron 
line is regarded as a signature of a standard optically thick disk, this 
suggests that such a disk is generically absent or truncated in LLAGNs.  
}

\item{
{\it Disklike H\al\ profiles.}\  Emission lines with broad,
double-peaked profiles, taken to be the kinematic signature of a
relativistically broadened disk, are found quite often in LLAGNs (Ho et al.
2000, and references therein; Shields et al. 2000; Barth et al.  2001; 
Eracleous \& Halpern 2001).  When fitted with a disk model, one infers that 
the disk has a relatively large inner radius ($\sim 10^3\, R_{\rm S}$).
}
\end{itemize}

\section{A Physical Picture of the Central Engine}

I propose that the above set of characteristics, common to most LLAGNs
studied in detail thus far, suggest that nearby galaxy bulges contain
central engines as schematically depicted in Figure~3.  Most galaxies
with bulges contain active nuclei because most, if not all, bulges contain
massive black holes.  This is consistent with the picture that has emerged from 
recent kinematical studies of nearby galaxies (e.g., Richstone 2004). In the
present-day Universe, and especially in the centers of big bulges, the amount
of gas available for accretion is quite small, plausibly well below the
Eddington rate for the associated black hole mass (Ho 2005).  In such a
regime, the low-density, tenuous material is optically thin and cannot cool 
efficiently.  Rather than settling into a classical optically thick, 
geometrically thin disk, the hot accretion flow assumes a quasi-spherical 
configuration, whose dynamics may be dominated by advection, convection, or 
outflows (see Quataert 2001 and Narayan, these proceedings.) 
For simplicity, I follow Quataert (2001) and simply call these
low-radiative efficiency accretion flows (LRAFs).  The existence of LRAFs
in these systems, or conversely the absence of classical thin disks extending
all the way to small radii (few $R_{\rm S}$), is suggested by their (1) low
luminosities, (2) low Eddington ratios, (3) low inferred radiative
efficiencies, (4) lack of a big blue bump, and (5) lack of relativistically
broadened Fe K\al\ lines.

%Figure 3
%%BoundingBox: 1 150 700 60
\begin{figure}
\vbox{
\hbox{
\hskip 0.0truein
\psfig{file=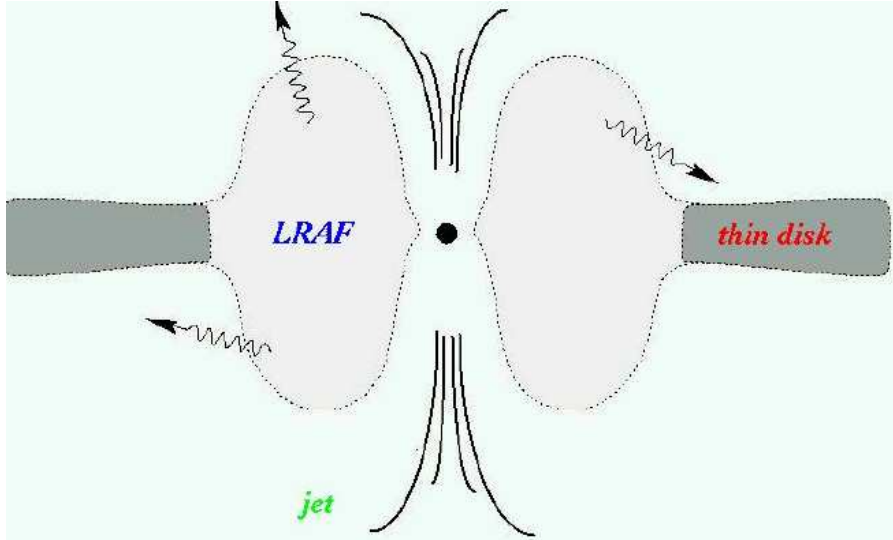,width=5.0truein,angle=0}
}
}
\caption{
A cartoon depicting the structure of the accretion flow
surrounding weakly active massive black holes.  An inner low-radiative
efficiency accretion flow (LRAF) irradiates an outer, truncated thin disk.
An additional compact jet component is needed.}
\end{figure}

Apart from a central LRAF, two additional components generally seem to be
required.  First, detailed considerations of the broad-band SED show that the
baseline LRAF spectrum underpredicts the observed radio power (e.g., Quataert
et al. 1999; Ulvestad \& Ho 2001).  Most of the radio luminosity, which is
substantial because these objects tend to be ``radio-loud,''  must come
from another component, and the most likely candidate is a compact jet.   Does
the puffed-up structure of an LRAF, or its propensity for outflows, somehow
facilitate the generation of relativistic jets?  Second, an  outer
thin disk, truncated at perhaps $\sim 100-1000\, R_{\rm S}$,  seems necessary to
explain (1) the existence of the IR excess in the SED (e.g., Quataert
et al. 1999) and (2) the prevalence of double-peaked broad emission lines
(Chen, Halpern, \& Filippenko 1989; Ho et al. 2000).  A large truncation radius
is also qualitatively consistent with the weakness or absence of broad Fe K\al\ 
emission.

Lastly, we note that low-ionization spectra may emerge quite naturally in the
scenario suggested above.  In the context of AGN photoionization models, it is
well known that LINER-like spectra can be produced largely by lowering the
``ionization parameter'' $U$, typically by a factor of $\sim 10$ below that in
Seyferts (e.g., Halpern \& Steiner 1983; Ferland \& Netzer 1983).  The
characteristically low luminosities of LINERs (Fig.~1{\it a}), coupled
with their low densities (Ho et al. 2003), naturally lead
to low values of $U$.  Two other effects, however, are also important in
boosting the low-ionization lines.  All else being equal, hardening the
ionizing spectrum (by removing the big blue bump) in photoionization
calculations creates a deeper partially ionized zone from which low-ionization
transitions, especially [O~I] \lamb\lamb 6300, 6363, are created.
Because of the prominence of the radio spectrum, cosmic-ray heating of the
line-emitting gas by the radio-emitting plasma may be nonnegligible; one
consequence of this process is again to enhance the low-ionization lines
(Ferland \& Mushotzky 1984).  Both of these effects should be investigated
quantitatively.

\acknowledgements
L. C. H. acknowledges financial support from the Carnegie Institution of
Washington and from NASA grants.

\theendnotes

\end{article}
\end{document}